\documentclass[12pt]{article}

\usepackage{newtxtext,newtxmath}

\usepackage{graphicx}

\usepackage[letterpaper,margin=1in]{geometry}

\renewenvironment{abstract}
	{\quotation}
	{\endquotation}

\date{}

\makeatletter
\renewcommand{\fnum@figure}{\textbf{Figure \thefigure}}
\renewcommand{\fnum@table}{\textbf{Table \thetable}}
\makeatother

\usepackage{scicite}

\usepackage{url}

\usepackage{soul}

\usepackage{bm}

\renewcommand{\vec}[1]{{\mathbf #1}}

\def\scititle{
	Experimental observation of three-dimensional Anderson localization of electromagnetic waves
}
\title{\bfseries \boldmath \scititle}

\author{
    Antton Goïcoechea$^{1\ast}$,
    Alexey Yamilov$^{2}$,
    Clément Ferise$^{1,3}$,\and
    Sergey E. Skipetrov$^{4}$, 
    Hui Cao$^{5}$, 
    Matthieu Davy$^{1\ast}$\and	
	\small{$^{1}$Université de Rennes, CNRS, IETR; 35000 Rennes, France}\and	
	\small{$^{2}$Physics Department, Missouri University of Science \& Technology; Rolla, Missouri 65409}\and	
    \small{$^{3}$Laboratory of Wave Engineering, {\'E}cole Polytechnique F{\'e}d{\'e}rale de Lausanne (EPFL), 1015 Lausanne, Switzerland}\and
	\small{$^{4}$Universit\'{e} Grenoble Alpes, CNRS, LPMMC; 38000 Grenoble, France}\and	
	\small{$^{5}$Department of Applied Physics, Yale University; New Haven, Connecticut 06520}\and
	\small$^\ast$Corresponding authors. E-mail: antton.goicoechea@dalembert.upmc.fr (A.G.)\and
    \small{matthieu.davy@univ-rennes.fr (M.D.)}
}

\begin{document} 

\maketitle


\begin{abstract} \bfseries \boldmath
A prominent phenomenon in contemporary condensed matter physics is Anderson localization---suppression of wave propagation in disordered systems as a result of interference effects. 
{Despite being observed with various types of waves over the years, all prior attempts to reach Anderson localization of light in three-dimensional systems have been hampered by experimental artifacts.}
Here, we report an unambiguous experimental proof of three-dimensional Anderson localization of microwaves in disordered metal aggregates.
{By studying samples with different metal volume fractions, we show a clear difference between diffusive and localized behaviors, and the latter is confirmed by a scaling analysis of transmitted beam width in excellent agreement with theoretical and numerical results.}
Our demonstration opens avenues for both fundamental studies and practical applications of this extraordinary phenomenon.
\end{abstract}


\noindent
Sixty-eight years ago, Anderson predicted the breakdown of electric conductance in disordered metals due to the interference of multiply scattered electronic wave functions~\cite{Anderson1958}. Later on, this phenomenon has been shown to extend beyond the realm of electronics \cite{john84,anderson85} and observed in a wide range of experiments with electrons~\cite{Kramer1993, imadaMetalinsulatorTransitions1998} and atoms~\cite{billyDirectObservationAnderson2008,jendrzejewskiThreedimensionalLocalizationUltracold2012} at low temperatures, water~\cite{guazzelliLocalizationShallowWater1983}, electromagnetic~\cite{chabanovStatisticalSignaturesPhoton2000, schwartzTransportAndersonLocalization2007, segevAndersonLocalizationLight2013}, and acoustic~\cite{weaverAndersonLocalizationUltrasound1990,Hu2008,Cobus2018,delmotte3DAndersonLocalization2026} waves. Anderson localization sets in when the size $L$ of a disordered medium exceeds the localization length $\xi$, which is determined by the disorder strength and the dimensionality of space.
In one- and two-dimensional fully disordered media without broken symmetries (orthogonal symmetry class), $\xi$ is finite no matter how weak the scattering is~\cite{Abrahams1979}, thus Anderson localization always takes place in a sufficiently large sample ($L >\xi$). 
In contrast, in three dimensions (3D), $\xi$ is infinite for weak disorder, and may become finite only when the product of wave number $k$ and scattering mean free path $\ell$ is of order unity---a strong disorder condition known as Ioffe-Regel criterion~\cite{ioffeNoncrystallineAmorphousLiquid1960}.
Strongly scattering dielectric powders (white paint) have been identified as promising media to realize Anderson localization of light in 3D, but all the experimental efforts in the past three decades have not provided any conclusive evidence~\cite{Wiersma1997, Scheffold1999, Wiersma1999, storzerObservationCriticalRegime2006, Sperling2013, scheffoldInelasticScatteringPuts2013}. 

Recently, the long-standing debate about the existence of Anderson localization of light in 3D has taken a dramatic turn~\cite{sperlingCan3DLight2016, Skipetrov2016}. Numerical simulations of large disordered systems have shown that 3D localization cannot be achieved in uncorrelated disordered dielectric media~\cite{yamilovAndersonLocalizationElectromagnetic2023}, presumably due to coupling between scatterers via longitudinal fields~\cite{Skipetrov2014, rezvaninaraghiPhaseTransitionsDiffusion2016, vantiggelenLongitudinalModesDiffusion2021}. However, the simulations have revealed the possibility of localizing light in 3D with metallic scatterers~\cite{yamilovAndersonLocalizationElectromagnetic2023}, into which electromagnetic waves barely penetrate, thus suppressing the longitudinal-field coupling. A dense disordered arrangement of metallic scatterers may be viewed rather simplistically as a collection of random resonators (air voids) between which light can propagate via waveguides (air channels). Thus, the situation is close to that of elastic waves propagating in a ``mesoglass'' of aluminum spheres in vacuum~\cite{Hu2008} and approximately maps onto the original Anderson's model of a quantum particle hopping between sites (air voids in our case) with random energies (random resonance frequencies).
{Metallic systems, despite being the subject of intense research in plasmonics~\cite{shalaevElectromagneticPropertiesSmallparticle1996, sarychevAndersonLocalizationSurface1999, lalanneLightInteractionPhotonic2018}, attracted less attention in the past in the context of
{localization of light in 3D} because of strong absorption of visible light that mars signatures of the phenomenon}.
Although a reduction of the diffusion coefficient (a prerequisite to Anderson localization) was observed in {3D mixtures of aluminum and Teflon spheres}~\cite{genackObservationPhotonLocalization1991}, there has been no unambiguous experimental demonstration of Anderson localization of light in 3D.

In this article, we observe Anderson localization of microwaves in a 3D slab of randomly packed aluminum particles. Under a point source excitation, the transmitted beam stops spreading in space for a metal filling fraction above a critical value. This is a tell-tale sign of Anderson localization, because it reveals the halt of diffusive spreading of wave energy even in the presence of absorption~\cite{Hu2008, Cherroret2010, Cobus2018, delmotte3DAndersonLocalization2026}. Any inelastic or nonlinear effect is excluded from the experimental data. Further evidence is provided by an analysis of the scaling of saturated beam width with sample thickness compared with predictions of the self-consistent theory of localization {and confirmed in numerical simulations}.


\subsection*{Samples and experiments}

\begin{figure} 
	\centering
    \includegraphics[width=0.99\textwidth]{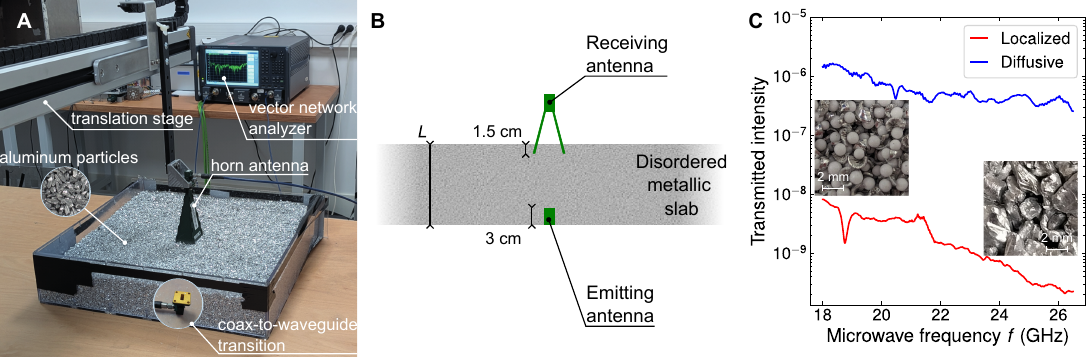}

	\caption{\textbf{Experimental setup and metal samples.}
		(\textbf{A}) Photograph of the experiment showing a {coax-to-waveguide} transition illuminating a slab made of randomly packed aluminum particles, and a horn antenna detecting the transmitted signals on the opposite side of the sample. The emitting antenna is placed inside the sample, and the horn antenna in reception is also plunged in the sample to avoid surface wave effects. {(\textbf{B}) Side view schematic of the experiment illustrating the emission 3~cm deep inside the sample of thickness $L$ and the measuring antenna {horn} plunging 1.5~cm inside the surface.} (\textbf{C}) Transmission spectra of two samples ($L=10$~cm): aluminum particles at metal filling fraction $\phi = $ 60\% (red curve, right inset) and aluminum flakes at $\phi = $ 15\% mixed with polystyrene spheres (blue curve, left inset). Insets are optical images of the two samples. Absence of pronounced structure in the transmission spectra reveals that the effects of structural order in our samples are negligible.}
	\label{fig:figure1} 
\end{figure}

Our samples consist of randomly packed aluminum particles of irregular shape, see Fig.~\ref{fig:figure1} and supplementary material~\cite{SM} for details. Their typical dimension of 2~mm is smaller than the wavelength of probing microwave ($\lambda_0 = 1.1$--1.7 cm) in the frequency range of $f = 18$--26 GHz. We estimate the aluminum filling fraction to be $\phi = 60\%$. The metallic particles are placed inside a plastic box of lateral dimension 45 cm $\times$ 45 cm and 12 cm height. Measurements are carried out for sample thicknesses $L$ from $6$ to $10$ cm.

The source is a rectangular coax-to-waveguide transition, with dimensions 0.4~cm $\times$ 1~cm, inserted at a depth $z' = 3$~cm inside the medium. On the reception side, a horn antenna with a rectangular aperture of 4~cm $\times$ 5~cm records the cross-polarized transmitted field at the back surface of the sample. The horn is inserted $1.5$~cm inside the sample{, which corresponds to approximately three times the localization length}
{as we will show later.} 
Positioning the emitting antenna inside the sample and inserting the receiving horn within the sample is essential to suppress the excitation of surface waves propagating along the sample interfaces.
Indeed, our numerical simulations detailed in the supplementary material~\cite{SM} show that external illumination of the sample leads to significant generation of surface waves at the interfaces of the samples.
These waves alter the spatial scaling of the transverse spreading of the transmitted beam, since the source aperture is virtually increased. The interplay between localized states and surface waves lies beyond the scope of the present work and will be addressed in future studies.

The transmission is recorded in the frequency domain using a narrowband spectral filter. Inelastic contributions are excluded by adjusting the filter bandwidth to about $1$~kHz (see details in supplementary material~\cite{SM}).
We obtain the spatially resolved spectrum of the transmitted field, $\psi(x, y, f)$, at $31\times31 = 961$ detector positions separated by 1~cm.
To improve statistical reliability, the field is measured ten times consecutively at each scanning position, with the horn being removed and re-inserted into the sample between measurements. Each insertion perturbs the surrounding particles, thereby generating a statistically independent configuration {at each measurement position}.

To demonstrate the {difference} between diffusion and localization, we repeat the same experiment for a sample with a lower metal filling fraction $\phi$. In this case, aluminum scatterers are 3 mm long, 1 mm wide and 0.1 mm thick flakes, obtained by machining of large aluminum cylinders. They are mixed with polystyrene spheres of refractive index 1.6 in order to reach $\phi = 15\%$, for which diffusive behavior is expected.

Spectra of average total transmitted intensity $T(f) = \langle \int |\psi(x, y, f)|^2 dxdy \rangle$, where $\langle\ldots\rangle$ denotes ensemble averaging, are shown in Fig.~\ref{fig:figure1}C for samples with thickness $L=10$~cm {(7~cm distance between emission and reception)} at low and high filling fractions. The spectra do not exhibit any pronounced structure, indicating negligible structural order, i.e., lack of correlations in positions of the aluminum scatterers, likely due to their irregular shapes and non-uniform sizes. Total transmission decreases with increasing metal filling fraction. $T(f)$ for $\phi = 60 \%$ is two orders of magnitude smaller than that for $\phi = 15 \%$.
{Due to the small thickness of the scatterers at $\phi = 15 \%$, the absorption is actually stronger than at $\phi = 60 \%$. The lower transmission of the $\phi = 60\%$ sample is therefore due to a significant increase in multiple scattering, which is crucial to reach the Anderson {localization} regime.}

\subsection*{Arrest of transverse spreading}
\begin{figure} 
	\centering
    \includegraphics[width=0.66\textwidth]{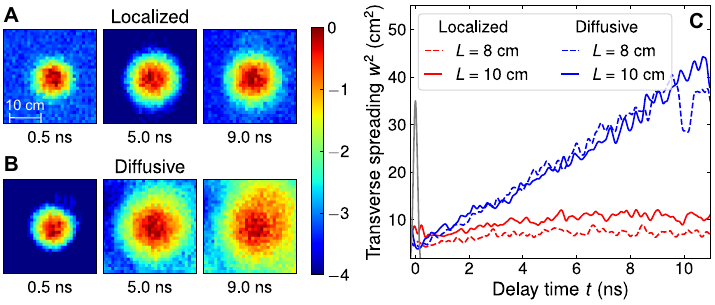}
	\caption{\textbf{Diffusion versus localization.}
		(\textbf{A}, \textbf{B}) Transmitted intensity profiles at the back surface of the sample (normalized and in logarithmic scale) recorded at three different times for the same samples as in Fig.~\ref{fig:figure1}, of metal filling fraction $\phi =$ 60\% (\textbf{A}) and 15\% (\textbf{B}).
        (\textbf{C}) Transmitted beam width squared $w^2(t)$ 
        in the diffusive ($\phi =$ 15\%) and localized ($\phi =$ 60\%) regimes for two sample thicknesses $L$ = 8 cm, 10 cm. A diffusion theory fit for $\phi =$ 15\% yields $D \simeq 0.85$ cm$^2$/ns. 
        The gray line shows the incident pulse of 0.25 ns full width at half maximum.}
	\label{fig:figure2} 
\end{figure}

We investigate the time-resolved transverse spreading of the transmitted beam following pulsed point-source excitation. The temporal evolution of the spatial field profile, $\psi(x, y,t)$, is obtained as the inverse Fourier transform of the spatially resolved transmission spectrum. Figures~\ref{fig:figure2}A and \ref{fig:figure2}B show snapshots of average transmitted intensity patterns $I(x, y, t) = \langle |\psi(x, y,t)|^2 \rangle$ at three representative times for the two samples with metal filling fractions $\phi = 15\%$ and $\phi = 60\%$, both of thickness $L=10$~cm. 
For $\phi = 15\%$, the transmitted wave spreads {across the entire output surface as time increases. This behavior is expected for diffusive transport of microwave, which is described by the same diffusion equation as the heat transfer. The energy density of multiply scattered waves satisfying a diffusion equation has been known for decades~\cite{ishimaruWavePropagationScattering1978}, and it is the deviations from this established transport regime that we are seeking as evidence for localization.}
In contrast {to the low filling fraction sample}, for $\phi = 60\%$ the transverse extent of the intensity profile becomes confined at long times. {This dramatically different behavior reveals that diffusion breaks down.} The temporal arrest of transverse spreading {actually} constitutes an unambiguous signature of Anderson localization
even in the presence of absorption~\cite{Hu2008,Cherroret2010}. While absorption reduces the overall transmitted intensity, it does not affect the transverse spreading dynamics.

The transverse spreading of the transmitted intensity profile, $I(x,y,t)$, is quantified by the time-dependent squared width of the transmitted beam, $w^2(t)$, which is related to the intensity participation ratio $\mathrm{PR}(t)$~\cite{yamilovAndersonLocalizationElectromagnetic2023}:
\begin{equation}
    w^2(t) = \frac{1}{2\pi} \mathrm{PR}(t) = \frac{1}{2\pi} \frac{\left[ \int I(x,y, t) dxdy\right]^2}{\int I(x,y, t)^2 dxdy}.
\end{equation}
This quantity is an important parameter to determine the transport regime and is similar to the position-dependent width used to demonstrate 3D Anderson localization of acoustic waves~\cite{Hu2008,Cobus2018,delmotte3DAndersonLocalization2026}. {Indeed, the time dependence of $w^2(t)$ directly relates to the spreading, or lack thereof, of the scattered wave intensity. It thus allows to discriminate between the different transport regimes discussed above.}

In the diffusive regime, {the intensity profile at the output surface of the medium is Gaussian: $I(x,y, t) \propto \exp [-(x^2+y^2)/4Dt]$, where $D$ is the photon diffusion coefficient. In this case,} $w^2(t)$ is independent of the sample thickness $L$ and increases linearly with time according to $w^2(t) = 4Dt$~\cite{Page1995}. For a filling fraction $\phi = 15\%$, Fig.~\ref{fig:figure2}C shows an identical linear growth of $w^2(t)$ for different sample thicknesses, $L = 8$~cm and $L = 10$~cm, yielding $D \simeq 0.85~\mathrm{cm}^2/\mathrm{ns}$. In contrast, Anderson localization leads to {a non-Gaussian intensity profile and} a saturation of $w^2(t)$ for sufficiently long times~\cite{Hu2008, Cobus2018, delmotte3DAndersonLocalization2026}, when $w^2(t)$ is approaching a thickness-dependent value $w_\infty^2$. This characteristic behavior is observed for $\phi =$ 60\%, as shown in Fig.~\ref{fig:figure2}C. 
A key advantage of our measurements in the frequency-domain with a narrow bandpass filter is the ability of excluding inelastic and nonlinear signals. When increasing the filter bandwidth from 500 Hz to 10 kHz, we observe no change in the transmitted beam width, thus eliminating possible contribution from fluorescence {which have plagued previous attempts to observe localization of light} ~\cite{Sperling2013,sperlingCan3DLight2016} (more details can be found in supplementary material~\cite{SM}).

\begin{figure} 
	\centering
    \includegraphics[width=0.99\textwidth]{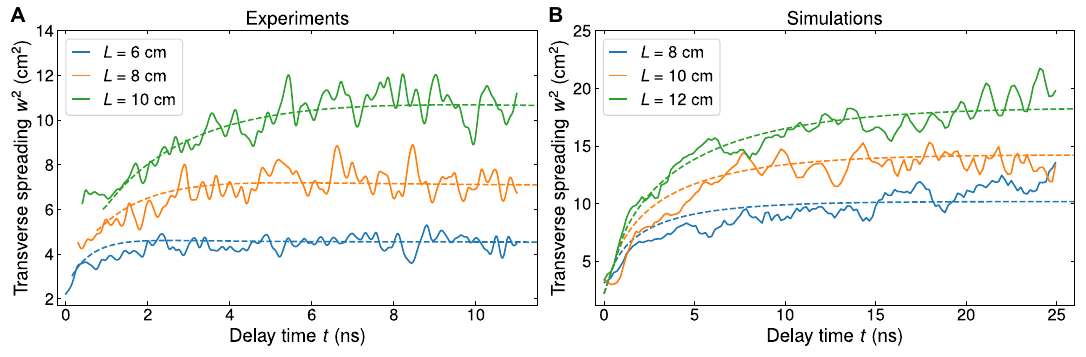}
	\caption{\textbf{Scaling analysis.}
		(\textbf{A}) Transmitted beam width squared $w^2(t)$ of the $\phi = 60\%$ sample along with the best fits obtained from
        the self-consistent theory
        (SCT) with the localization length {$\xi=$ 0.4~cm ($L=$ 6~cm), $\xi=$ 0.5~cm ($L=$ 8~cm), and $\xi=$ 0.6~cm ($L=$ 10~cm)}. (\textbf{B}) Transmitted beam width squared $w^2(t)$ for numerical simulations of randomly placed, overlapping PEC spheres with the same $\phi$, detailed in the supplementary material~\cite{SM}.
        SCT fit yields the localization length $\xi = $~1 cm.}
	\label{fig:figure3} 
\end{figure}

Unlike in previous demonstrations of 3D Anderson localization of acoustic waves \cite{Hu2008,Cobus2018,delmotte3DAndersonLocalization2026}, our experimental setup offers the unique opportunity of studying the scaling with tunable sample thickness. This flexibility allows us to further validate the localized regime through a scaling analysis of the squared beam width $w^2(t)$ as a function of the sample thickness $L$ for the $\phi = 60\%$ sample. Measurements are conducted for thicknesses ranging from $L = 6$ to $10$~cm, corresponding to source–detector separations between {3} and {7}~cm.
The experimental data are compared to predictions of the self-consistent theory (SCT) of Anderson localization, which accounts for interference effects by renormalizing the position-dependent diffusion coefficient based on the return probability of the wave \cite{wolfle2010,tiggelen2000,cherroret08}. As shown in Fig.~\ref{fig:figure3}A, SCT provides excellent fits to the experimental $w^2(t)$ with
the localization length $\xi$ as the only fit parameter \cite{SM}. The fits yield
$\xi$ ranging between {0.4} and 0.6~cm. These localization lengths are significantly shorter than the sample thicknesses, unambiguously confirming Anderson localization of microwaves in our samples.
Note that $w^2(t)$ does not vanish for $t \to 0$ due to the finite size of both emitting and receiving antennas that we model by convolving point-antenna results with a Gaussian profile of width $w_0 = 1.46$~cm in our SCT calculations~\cite{SM}.

We complement our experimental findings with numerical simulations of disordered samples composed of overlapping metallic spheres at a filling fraction $\phi = 60\%$ \cite{yamilovAndersonLocalizationElectromagnetic2023}. To minimize surface-wave contributions, the transmitted intensity is evaluated at a depth of 1.5~cm below the sample surface. The squared beam width $w^2(t)$ is computed by averaging over 10--20 independent disorder realizations.
As shown in Fig.~\ref{fig:figure3}B, the numerical results are in excellent agreement with SCT
for $\xi = 1$~cm. The value of $\xi$ depends on specific details of the microstructure, and hence is not expected to match the experimental case exactly. In addition, the spatial resolution of the numerical mesh is limited, preventing subwavelength features smaller than $\lambda_0/20 \simeq 0.7$~mm to be resolved.

\begin{figure} 
	\centering
    \includegraphics[width=0.66\textwidth]{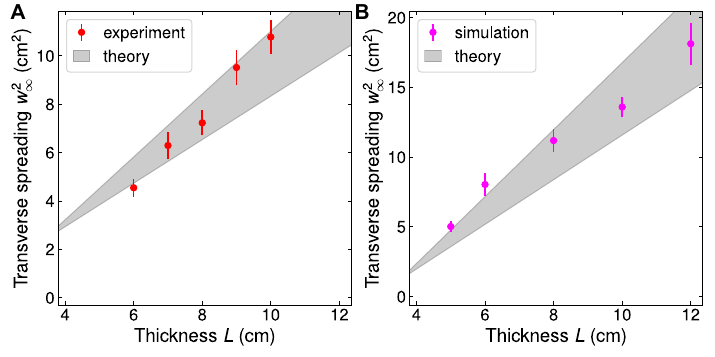}
	\caption{\textbf{Saturation of transmitted beam width in the long-time limit.}
		(\textbf{A}, \textbf{B}) Long-time limit of the transmitted beam width squared $w^2_{\infty}$ as a function of sample thickness $L$ obtained in experiments (\textbf{A}) and simulations (\textbf{B}). The red dots correspond to the experimental value (\textbf{A}), the magenta dots to the simulations (\textbf{B}). The gray shaded areas show SCT predictions given by Eq.~(\ref{eq:w2 inf}) for $0.45 <\xi<0.65$~cm (\textbf{A}), and by Eq.~(\ref{w2inftyinside}) in supplementary material~\cite{SM} for $0.8 <\xi<1.2$~cm (\textbf{B}). The predictions are not expected to be accurate for $L \lesssim 4$~cm.}
	\label{fig:figure4} 
\end{figure}

{In addition to capturing the temporal evolution of $w^2(t)$, SCT also yields its long-time saturation value $w^2_{\infty}$. For $L-z' \gg \xi$, SCT results for $w^2_{\infty}$ follow an approximate expression~\cite{SM}:
\begin{equation}
    w_\infty^2 \simeq w_0^2 + 2 (L-z') \xi \left[ 1 - \frac{\xi}{4(L-z')} \right].
	\label{eq:w2 inf}
\end{equation}
This prediction is compared with results of experiments in Fig.~\ref{fig:figure4}A. In simulations, the transmitted field is computed inside the sample and the corresponding $w_\infty^2$ is not given by Eq.~(\ref{eq:w2 inf}), but it can still be compared to SCT predictions~\cite{SM}, see Fig.~\ref{fig:figure4}B.
$w^2_{\infty}$ is obtained as a time average of $w^2(t)$ for $t>7$~ns in the experiment and for $t>15$~ns in the simulations, which corresponds, in both cases, to times at which the plateau of $w^2(t)$ is reached. In both the experiment and simulations, $w^2_{\infty}$ exhibit slight fluctuations, but remain consistent with {$\xi = 0.55 \pm 0.1$~cm} and {$\xi = 1 \pm 0.2$~cm}, respectively, for all $L$.}
{This scaling analysis shows that SCT is able to capture the dependence of the beam spreading on $L$ and confirms that Anderson localization regime has been reached.}

\subsection*{Conclusion}
In conclusion, we demonstrate Anderson localization of electromagnetic waves in a 3D disordered medium, thus settling the long-standing debate about its existence. Since our 3D disordered samples are easy to fabricate, a fine tuning of metal filling fraction and sample thickness may enable a rigorous finite-size scaling analysis and experimental determination of the critical exponent of Anderson localization transition in the future. Extension of our results to the optical spectrum, including visible and near-infrared, may lead to promising applications of Anderson localization in photonics technologies such as photo-catalysis, optical sensing, energy conversion and storage.

\newpage



\clearpage 

%
\bibliography{AndersonLoc_EM3D} 
\bibliographystyle{sciencemag}

%
%
%
%
%
%


\section*{Acknowledgments}
{In memory of Bart Van Tiggelen, whose pioneering spirit continues to inspire our work.} We thank Laurent Cronier for his help in preparing the samples.
\paragraph*{Funding:}
This work is supported by the European Union through European Regional Development Fund (ERDF), Ministry of Higher Education and Research, CNRS, Brittany region, Conseils Départementaux d’Ille-et-Vilaine and Côtes d’Armor, Rennes Métropole, and Lannion Trégor Communauté, through the CPER Project CyMoCod. C.F. acknowledges funding from the French ‘Ministère de la Défense, Direction Générale de l’Armement’. M.D. acknowledges the Institut Universitaire de France. A.Y. is supported by the US National Science Foundation under grant no. DMR-1905442. H.C. acknowledges funding support of the US National Science Foundation under grant no. DMR-1905465. The authors sincerely thank Professor Zongfu Yu and
Flexcompute Inc. for providing us access to the Tidy3D
software for running the FDTD numerical simulations
described in this work.
\paragraph*{Author contributions:}\mbox{} \\
Conceptualization: AG, AY, SES, HC, MD\\
Formal analysis: AG, AY, SES, HC, MD\\
Funding acquisition: MD\\
Investigation: AG, CF, AY, SES, HC, MD\\
Methodology: AG, AY, SES, HC, MD\\
Software: AG, AY, SES\\
Validation: AG, CF, AY, SES, HC, MD\\
Visualization: AG, MD\\
Writing – original draft: AG, MD\\
Writing – review \& editing: AG, CF, AY, SES, HC, MD\\
\paragraph*{Competing interests:}
There are no competing interests to declare.
\paragraph*{Data and materials availability:}
The data and codes to generate the figures of this study are available at~\cite{data}.

\subsection*{Supplementary materials}
Materials and Methods\\
Figs. S1 to S5\\
References \textit{(35-\arabic{enumiv})} 


\newpage


\renewcommand{\thefigure}{S\arabic{figure}}
\renewcommand{\thetable}{S\arabic{table}}
\renewcommand{\theequation}{S\arabic{equation}}
\renewcommand{\thepage}{S\arabic{page}}
\setcounter{figure}{0}
\setcounter{table}{0}
\setcounter{equation}{0}
\setcounter{page}{1} 


\begin{center}
\section*{Supplementary Material for\\ \scititle}

Antton Goïcoechea$^{1\ast}$,
Alexey Yamilov$^{2}$,
Clément Ferise$^{1,3}$,\\
Sergey E. Skipetrov$^{4}$, 
Hui Cao$^{5}$, 
Matthieu Davy$^{1\ast}$\\	
\small{$^{1}$Université de Rennes, CNRS, IETR; 35000 Rennes, France}\\	
\small{$^{2}$Physics Department, Missouri University of Science \& Technology;}{Rolla, Missouri 65409}\\	
\small{$^{3}$Laboratory of Wave Engineering, {\'E}cole Polytechnique F{\'e}d{\'e}rale de Lausanne (EPFL),}{1015 Lausanne, Switzerland}\\
\small{$^{4}$Universit\'{e} Grenoble Alpes, CNRS, LPMMC; 38000 Grenoble, France}\\	
\small{$^{5}$Department of Applied Physics, Yale University; New Haven, Connecticut 06520}\\
\small$^\ast$Corresponding authors. E-mail: antton.goicoechea@dalembert.upmc.fr (A.G.)\\
\small{matthieu.davy@univ-rennes.fr (M.D.)}
\end{center}

\subsubsection*{This PDF file includes:}
Materials and Methods\\
Figures S1 to S5

\newpage


\subsection*{Materials and Methods}

\subsubsection*{Experiments}
The high filling fraction $\phi =$ 60\% sample is made of aluminum particles of irregular shape, with a typical ``radius" of 2 mm. At microwave frequencies, aluminum behaves as a conductor {with a penetration depth of $\sim 0.1$~µm.}
The low filling fraction $\phi =$ 15\% sample is made of aluminum flakes mixed with polystyrene spheres of radius 1 mm. The flakes are obtained by machining aluminum cylinders in the mechanical workshop of IETR. They have irregular shape and strong curvature. The polystyrene spheres are added to reduce the volume fraction $\phi$ of aluminum flakes, but themselves have little impact on electromagnetic wave transport due to small refractive index of polystyrene $n\sim 1.6$ {compared to aluminum} in the microwave range of our experiment~\cite{jofreWirelessMicrowaveSensing2021, palaciosSuperheterodyneMicrowaveSystem2022}. $\phi$ is determined by weighing the samples.
{Changing the thickness $L$ of the sample in a consistent way was done by adding a given weight of particles. This provided a simple way to always add the same quantity of particles in the system and only required flattening the surface afterward.}

The source and the probe operate in the K-band (18-26 GHz).
The source is a rectangular coax-to-waveguide transition, with dimensions 0.4~cm $\times$ 1~cm, inserted 3~cm inside the medium. On the reception side, a horn antenna with a rectangular aperture of 4~cm $\times$ 5~cm records the cross-polarized transmitted field at the back surface of the sample. 
Horn antennas are directive probes that make it possible to measure transmission even through strongly scattering systems, but at the cost of an integration of the field over an effective aperture of $A \sim \lambda_0^2$ in this case. We measure the field transmission coefficients $\psi(x,y;f)$ between 18 and 26 GHz using a Vector Network Analyzer. 
The receiving horn is inserted $1.5$~cm inside the sample and is translated over a $31\times31$ grid of points using a motorized translation stage. The spacing between two points is 1~cm. The temporal variation of the field $\psi(x,y;t)$ is then obtained from an inverse Fourier transform of the spectrum of the field transmission coefficient, for an incident Gaussian pulse of a center frequency $f_0 = 22.25$ GHz and a width of 1.5 GHz.

Because the spectra of transmission coefficients are measured for individual samples, the averaging over an ensemble of independent configurations denoted by $\langle \ldots \rangle$ in the statistical analysis of transmitted intensity is performed spatially over the output surface.

\begin{figure} 
	\centering
    \includegraphics[width=0.66\textwidth]{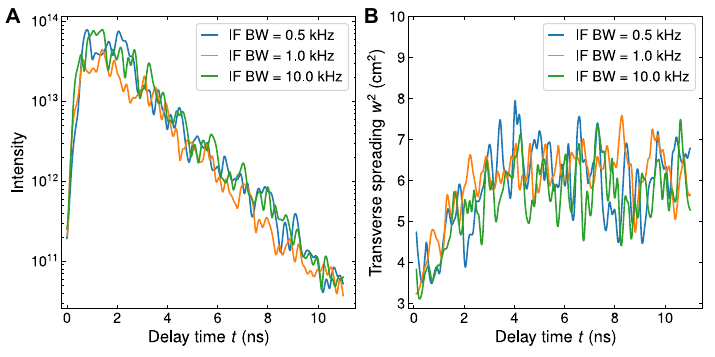}

	\caption{{\textbf{Microwave transport with varying detection bandwidth.}
		(\textbf{A}) Intensity of transmitted microwaves, detected with the intermediate frequency (IF) filter bandwidth of 0.5 kHz, 1 kHz, 10 kHz, showing the same exponential decay in time. The sample thickness is $L$ = 7 cm, the metal filling fraction is $\phi$ = 60\%.  (\textbf{B}) Transverse width squared $w^2(t)$ of transmitted beam through the same sample as in A, for the three IF bandwidths. The transverse spreading of transmitted beam in time is nearly identical for the three detection bandwidths, confirming that contributions from broadband background and inelastic signals like fluorescence are negligible.}}
	\label{fig:IF_BW} 
\end{figure}

To exclude broadband inelastic and nonlinear signals such as fluorescence, the transmitted microwave is measured in the frequency domain by a Vector Network Analyzer with narrowband filters of 500 Hz, 1 kHz and then 10 kHz for a sample of $L=$ 7~cm and $\phi=$ 60\%. Figure~\ref{fig:IF_BW}A shows the time-resolved intensity of transmitted microwaves for filter bandwidths of 500 Hz, 1 kHz, and 10 kHz.
The normalized traces exhibit the same decay in time. In Fig.~\ref{fig:IF_BW}B, transverse spreading of the transmitted beam is identical for the three bandwidths. Furthermore, a non-monotonic variation of the transverse beam width with time, which could be attributed to fluorescence based on previous optical experiments~\cite{sperlingCan3DLight2016}, is not observed.



\subsubsection*{Self-consistent theory of localization}
Consider the intensity Green's function $C(\vec{r},\vec{r}',t)$, equal to the average intensity of a wave at position $\vec{r}$ and time $t$ after emission of an infinitely short pulse by a point source at $\vec{r}'$ at $t' = 0$. According to the self-consistent theory (SCT) of localization, its Fourier transform $C(\vec{r},\vec{r}',\Omega)$ obeys \cite{wolfle2010,tiggelen2000,cherroret08}
\begin{equation}
\left[-i \Omega -\nabla \cdot D(\vec{r},\Omega) \nabla \right] C(\vec{r},\vec{r}',\Omega) = \delta(\vec{r}-\vec{r}')
\label{sceq1}
\end{equation}
where $D(\vec{r},\Omega)$ is a position-dependent diffusion coefficient to be determined self-consistently:
\begin{equation}
\frac{1}{D(\vec{r},\Omega)} = \frac{1}{D_B} + \frac{12\pi}{k^2 \ell} C(\vec{r},\vec{r},\Omega)
\label{sceq2}
\end{equation}
$D_B = c\ell/3$ is the bare value of $D$ in the absence of localization effects, $c$ is the speed of the wave (light here), $k$ is the wave number, $\ell$ is the scattering mean free path. It is, in principle,  possible to differentiate between the bare transport mean free path $\ell_B^*$ entering into the expression of $D_B$ and $\ell$, but here we adopt the simplest version of SCT with $\ell_B^* = \ell$. 

We assume that the mobility edge is determined by the Ioffe-Regel criterion $k\ell = 1$ \cite{shengIntroductionWaveScattering2006}. When $k\ell \gg 1$, Eq.\ (\ref{sceq2}) yields $D(\vec{r},\Omega) = D_B$ and Eq.\ (\ref{sceq1}) reduces to the standard diffusion equation of the transport theory. In contrast, for $k\ell < 1$ Eqs.\ (\ref{sceq1}) and  (\ref{sceq2}) yield a different behavior corresponding to Anderson localization. In the infinite medium, Eqs.\ (\ref{sceq1}) and (\ref{sceq2}) predict that for $k\ell < 1$ and in the long-time limit, the intensity distribution due to a short pulse emitted at $\vec{r}'$ at time $t' = 0$ is \cite{Cobus2018}
 \begin{equation}
C(\vec{r},\vec{r}', t \to \infty) = \frac{1}{4\pi\xi^2 |\vec{r}-\vec{r}'|}
\exp\left( -\frac{|\vec{r}-\vec{r}'|}{\xi} \right)
\label{infpulse}
\end{equation}
where the localization length is
\begin{equation}
\xi = 6 \ell \frac{(k \ell)^2}{1-(k \ell)^4}
\label{xi}
\end{equation}

In a slab of thickness $L$ and area $A \gg L^2$ confined between planes $z=0$ and $z=L$, $D(\vec{r},\Omega) = D(z,\Omega)$ and Eqs.\ (\ref{sceq1}) and (\ref{sceq2}) should be complemented  with boundary conditions \cite{haskell94,cherroret08}
\begin{equation}
\left[ C(\vec{r},\vec{r}', \Omega) \pm z_0 \frac{D(z, \Omega)}{D_B} \frac{\partial}{\partial z} C(\vec{r},\vec{r}', \Omega) \right]_{z=0,L} = 0 
\label{bc}
\end{equation}
where $z_0$ is the extrapolation length depending on the internal reflections at the sample boundary.  Assuming no internal reflections yields $z_0 = 2\ell/3$. It is convenient to work with the spatial Fourier transform of $C$ with respect to $\boldsymbol{\rho} = (x, y)$: $C(\vec{q}_{\perp}, z, z', \Omega)$. The stationary transmission of a plane wave through the slab is found as
\begin{equation}
\begin{aligned}
T  &= -D(z,  \Omega=0)
\left. \frac{\partial}{\partial z} 
C(\vec{q}_{\perp} = 0, z, z' = \ell, \Omega=0)
\right|_{z=L}
\propto \exp(-L/\xi)
\end{aligned}
\label{trans}
\end{equation}

Equations (\ref{sceq1}) and (\ref{sceq2}) also allow for computing the intensity profile $T(\boldsymbol{\rho},t)$ in transmission of a short pulse emitted by a source located at $\vec{r}' = \{\boldsymbol{0}, z' \}$ through a slab of disordered medium of thickness $L$, as described in Refs.\ \cite{Cherroret2010,Cobus2018}, for $z' = \ell$ corresponding to a wave incident on the slab from outside. Assuming that finite sizes of both source and detector can be accounted for by a convolution with a Gaussian $\propto \exp(-\rho^2/w_0^2)$, we have
\begin{equation}
\begin{aligned}
T(\vec{q}_{\perp},  \Omega)  &= \left. -D \frac{\partial}{\partial z} 
C(\vec{q}_{\perp}, z, z', \Omega) \right|_{z=L}
\\
T(\boldsymbol{\rho}, t) &= \int\limits_{-\infty}^{\infty} \frac{d\Omega}{2\pi} e^{-i \Omega t} 
\int \frac{d^2\vec{q}_{\perp}}{(2\pi)^2} e^{-i \vec{q}_{\perp} \boldsymbol{\rho}}
e^{-\vec{q}_{\perp}^2 w_0^2/4}
T(\vec{q}_{\perp}, \Omega)
\end{aligned}
\label{trans2}
\end{equation}

\noindent Average intensity of a wave in an arbitrary plane $z = \text{const}$ inside the slab is simply 
\begin{equation}
\begin{aligned}
C(\boldsymbol{\rho}, z, z', t) &= \int\limits_{-\infty}^{\infty} \frac{d\Omega}{2\pi} e^{-i \Omega t} 
\int \frac{d^2\vec{q}_{\perp}}{(2\pi)^2} e^{-i \vec{q}_{\perp} \boldsymbol{\rho}}
e^{-\vec{q}_{\perp}^2 w_0^2/4}
C(\vec{q}_{\perp}, z, z', \Omega)
\end{aligned}
\label{intensity}
\end{equation}

\noindent The spatial extension of $T$ or $C$ in a plane $z = \text{const}$ can be characterized by a participation ratio
\begin{eqnarray}
\text{PR}(t) = \frac{\left[ \int d^2 \boldsymbol{\rho} T(\boldsymbol{\rho},t) \right]^2}{\int d^2 \boldsymbol{\rho} T(\boldsymbol{\rho},t)^2}
\label{pr}
\end{eqnarray}
(and similarly for $C$) or by an effective width of the transmission or intensity profile
\begin{eqnarray}
w(t)^2 = \frac{1}{2\pi} \text{PR}(t)
\label{w2}
\end{eqnarray}
defined such that $w(t)^2 = \sigma(t)^2$ for a Gaussian profile $T(\boldsymbol{\rho},t) \propto \exp(-\rho^2/\sigma(t)^2)$.

The simplest version of SC theory described above has a single free parameter, the localization length $\xi$, which determines $\ell$ via Eq.\ (\ref{xi}), $k = 2\pi f/c$ being fixed by the frequency $f$. We solve Fourier transforms of Eqs.\ (\ref{sceq1}), (\ref{sceq2}) and (\ref{bc}) for $C(\vec{q}_{\perp}, z, z', \Omega)$ for $f$ equal to the central frequency $f_c = 22$ GHz of the emitted pulse numerically by discretizing $z$ and $q$ in sufficiently small steps \cite{Cobus2018}. The width of the intensity profile in transmission or in a plane $z = \text{const}$ inside the sample  is then calculated using Eqs.\ (\ref{trans2}--\ref{w2}).

\begin{figure} 
	\centering
    \includegraphics[width=0.6\textwidth]{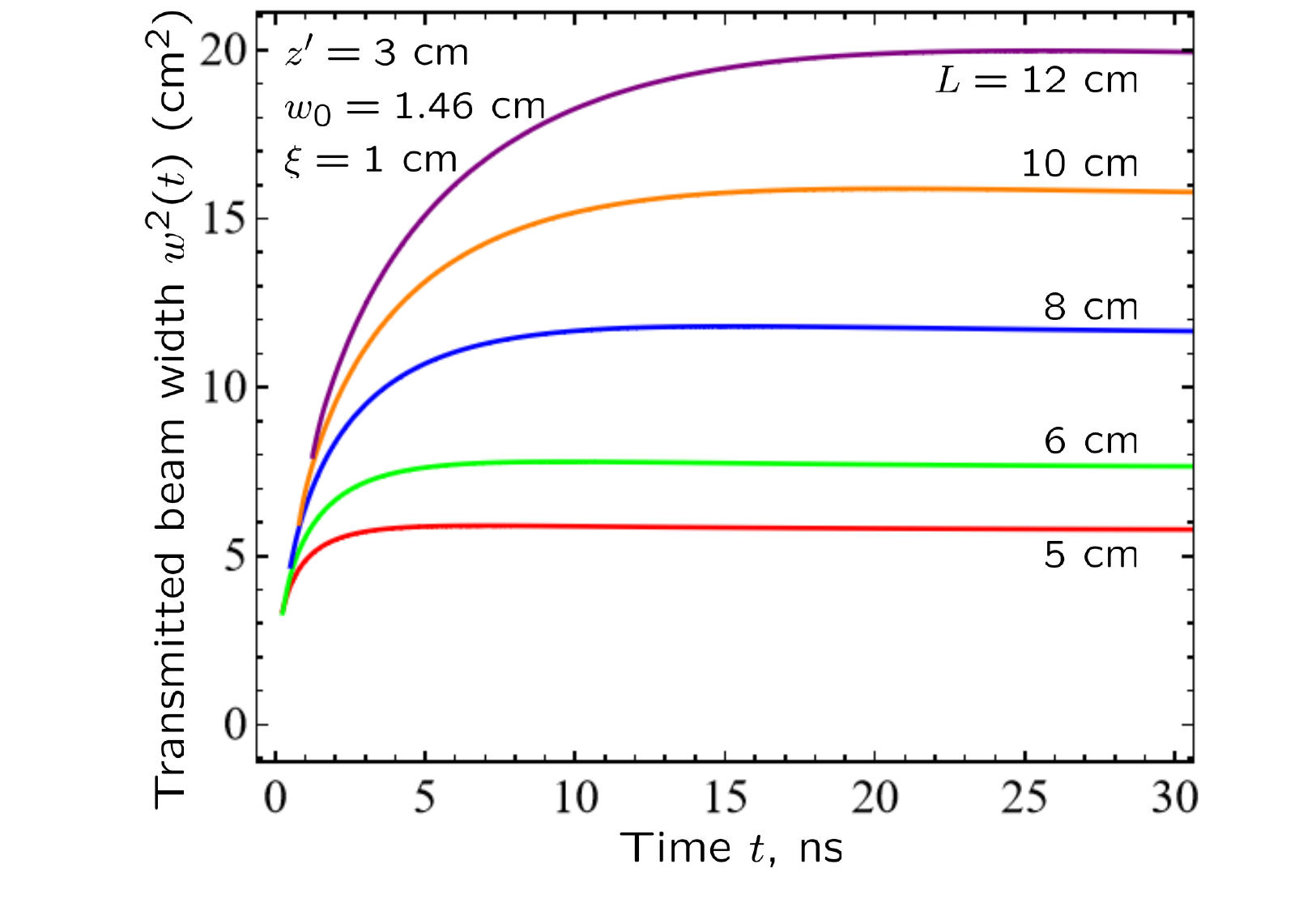}
	\caption{{\textbf{Theoretical prediction of transverse spreading of the transmitted beam.}}
		Square of the width of the transmitted intensity profile $T(\boldsymbol{\rho},t)$ for  $z' = 3$ cm, $w_0 = 1.46$ cm, $\xi = 1$ cm and several values of slab thickness $L$.}
	\label{fig_w2_sc} 
\end{figure}

\begin{figure} 
	\centering
    \hspace*{-8mm}
    \includegraphics[width=0.54\textwidth]{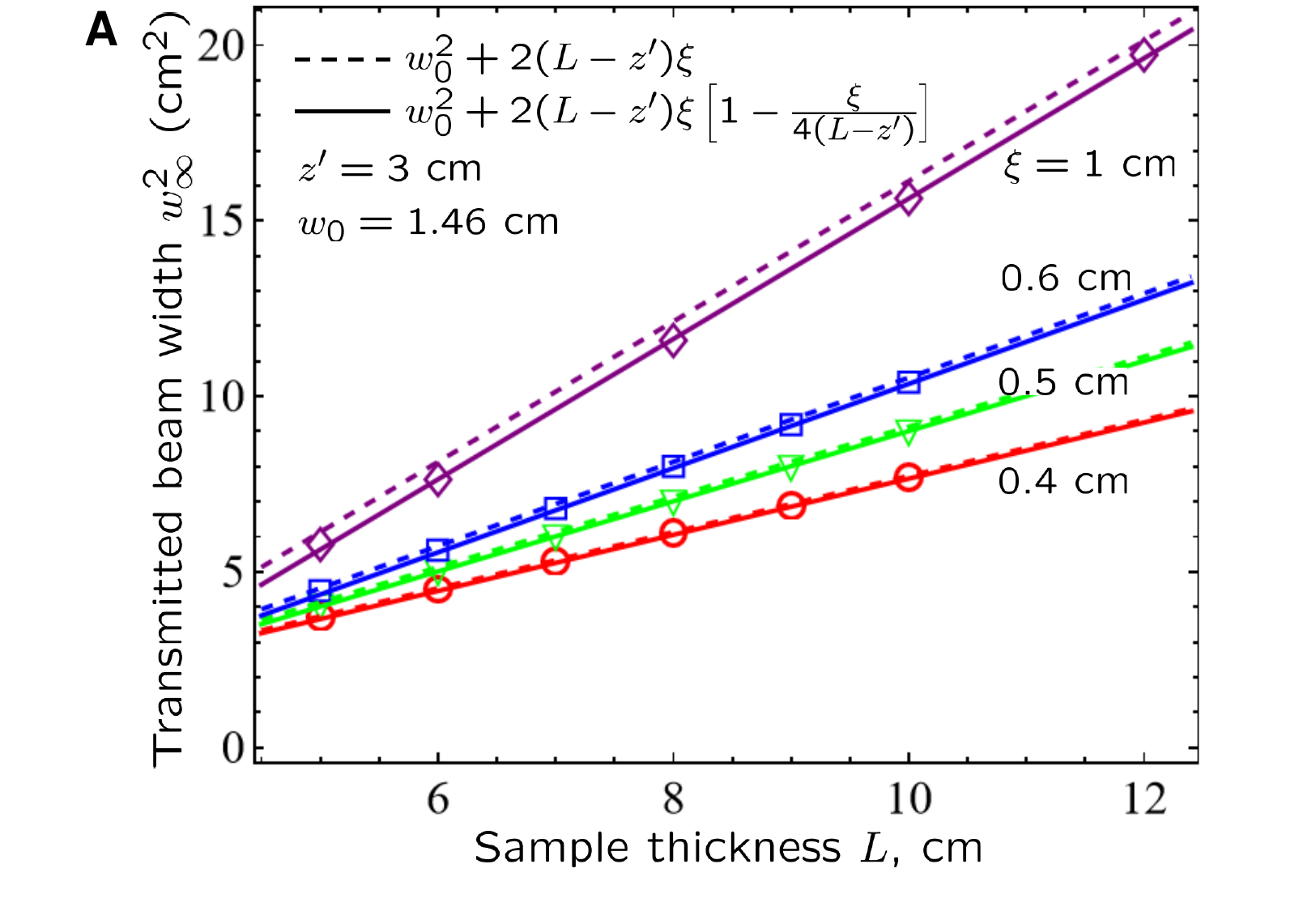}
    \hspace*{-8mm}
\includegraphics[width=0.54\textwidth]{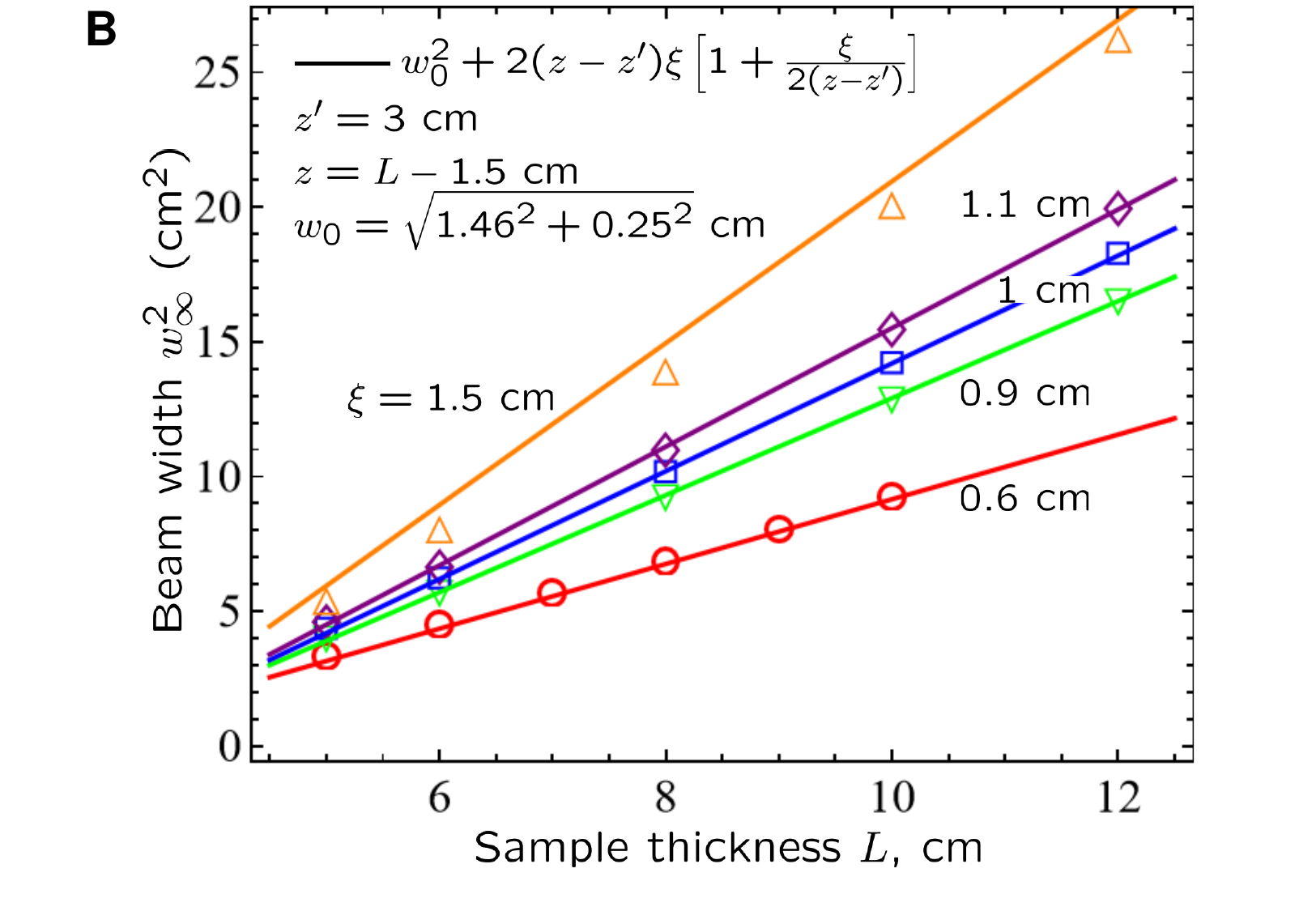}
	\caption{{\textbf{Saturation level of the width in transmission and inside the sample.}}
		\textbf{(A)} Saturated square of the transverse width of the transmitted beam $w(t)^2$ in the limit of long time $t \to \infty$ for $z' = 3$ cm, $w_0 = 1.46$ cm, and several values of $\xi = 0.4$--1 cm, as a function of slab thickness $L$. Symbols show results of numerical solution of Eqs.\ (\ref{sceq1}), (\ref{sceq2}) and (\ref{bc}). Dashed lines show the infinite-medium approximation, solid lines provide a better approximation (\ref{w2infty}) in the considered ranges of $\xi$ and $L$.
        \textbf{(B)} The same as \textbf{(A)} but inside the sample, at a depth $z = L-1.5$ cm, for five values of $\xi = 0.6$--1.5 cm.
        Equation\ (\ref{w2inftyinside}) shown by solid lines provides a good approximation to numerical results for $z'$, $z-z'$, $L-z \gg \xi$, but start to deviate from the latter for the largest value of $\xi = 1.5$ cm (orange line and symbols). $w_0$ takes into account the effective sizes of both the emitting (0.25 cm) and receiving (1.46 cm) antennas. }
	\label{fig_saturation_exp} 
\end{figure}

We show typical results following from SCT for the width squared of the intensity profile in transmission in Figs.\ \ref{fig_w2_sc} and \ref{fig_saturation_exp}A. It is quite remarkable that for $L-z' \gg \xi$, a good approximation to the value $w_{\infty}^2$ at which $w(t)^2$ saturates at long times, can be obtained by neglecting the boundary conditions (\ref{bc}) and solving Eqs.\ (\ref{sceq1}) and (\ref{sceq2}) in the infinite medium [see Eq.\ (\ref{infpulse})], which yields $w_{\infty}^2 \simeq w_0^2 +  2 (L-z') \xi$ (dashed lines in Fig.\ \ref{fig_saturation_exp}A). An even better approximation to numerical results is provided by the following approximate expression:
\begin{eqnarray}
w_{\infty}^2 \simeq w_0^2 + 2 (L-z') \xi \left[ 1 - \frac{\xi}{4(L-z')} \right]
\label{w2infty}
\end{eqnarray}
This expression is shown in Fig.\ \ref{fig_saturation_exp}A by solid lines. 

In FDTD simulations reported in the main text, $w(t)^2$ is evaluated at a depth $z = L-1.5$ cm inside the sample in order to reduce the impact of surface waves (see also the explanations in the section ``Surface wave suppression'' below). SCT also allows for a comparison with these results by using Eq.\ (\ref{pr}) with $T(\bm{\rho},t)$ replaced by $C(\vec{r} = \{\bm{\rho}, z \}, \vec{r}' = \{0, z' \}, t)$. Typical results for $w_{\infty}^2$ are illustrated in Fig.\ \ref{fig_saturation_exp}B. In the considered range of parameters, a good approximation to SCT results is provided by a simple formula
\begin{eqnarray}
w_{\infty}^2 \simeq w_0^2 + 2 (z-z') \xi \left[ 1 + \frac{\xi}{2(z-z')} \right]
\label{w2inftyinside}
\end{eqnarray}
shown in Fig.\ \ref{fig_saturation_exp}B by solid lines.

\subsubsection*{Finite-difference time-domain simulations\label{sec:numerical_method}}

\begin{figure} 
	\centering
    \includegraphics[width=0.5\textwidth]{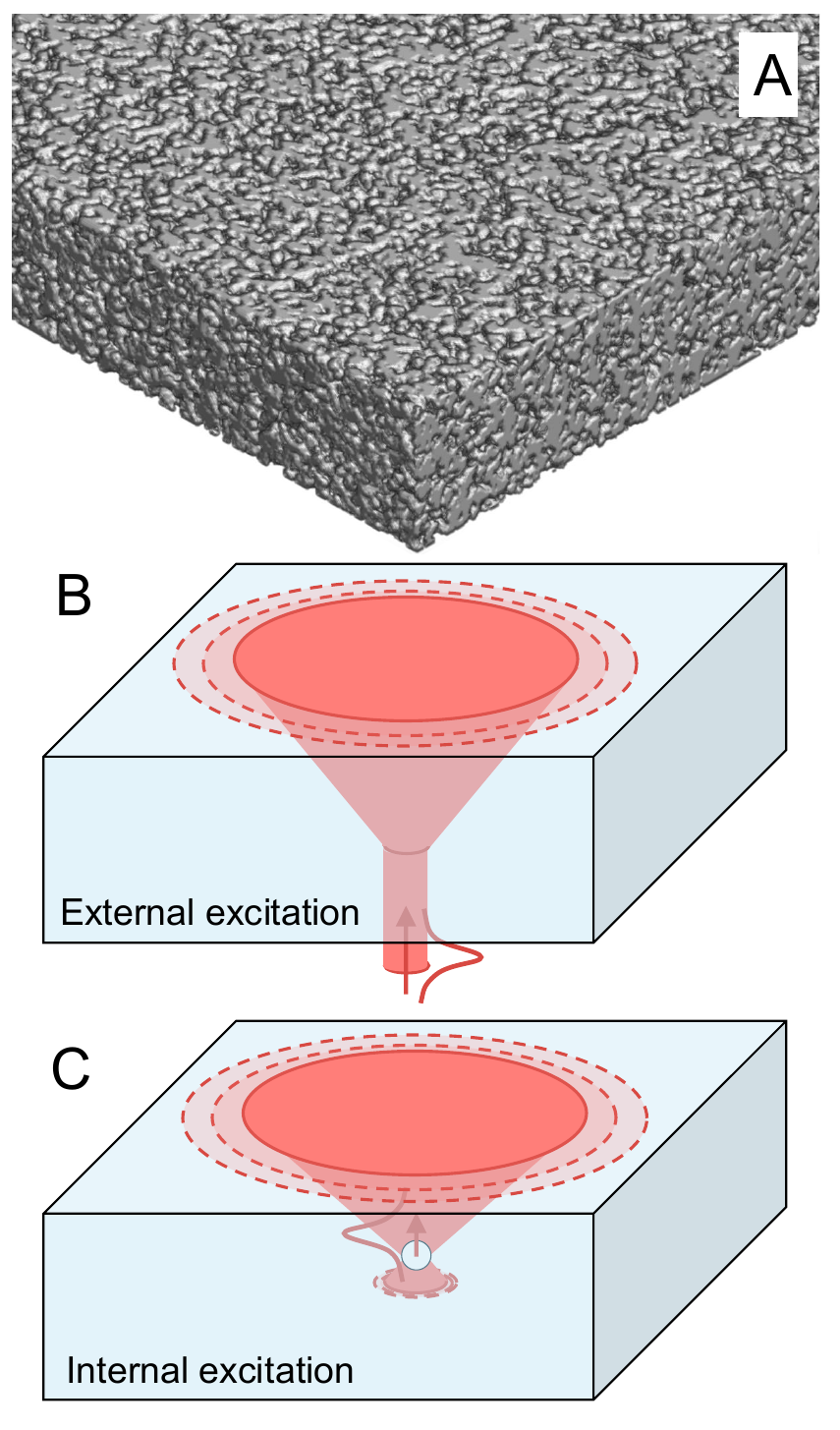}
	\caption{\textbf{Simulation geometry.}
		\textbf{(A)} Representative disordered slab consisting of randomly positioned overlapping PEC spheres of radius $r = $ 0.185~cm at volume filling fraction $\phi =$ 60\%.
        \textbf{(B)} External-excitation geometry: a small (approximately half-wavelength) Gaussian source placed in air illuminates the bottom face of the slab. 
        This configuration strongly excites surface waves at the air–composite interfaces and is shown for comparison. 
        \textbf{(C)} Internal-excitation geometry used throughout this work: the source, approximately half-wavelenth in size, is embedded 3~cm inside the medium within a small air void, and the detection plane is located inside the sample, 1.5~cm from the top sample face. 
        This arrangement probes bulk transport and suppresses coupling to interfacial surface waves, cf. Fig.~\ref{fig:surface_waves}.}
	\label{fig:simulation_geometry} 
\end{figure}

Numerical simulations employ the hardware-accelerated finite-difference time-domain (FDTD) method described in Ref.~\cite{yamilovAndersonLocalizationElectromagnetic2023}. 
With center frequency $f_0 = 22$~GHz ($\lambda_0 = 1.36$~cm), we simulate 3D slabs of transverse dimensions $45~\text{cm} \times 45~\text{cm}$ and thicknesses $L = 5, 6, 8, 10, 12$~cm, filled with randomly positioned, overlapping PEC spheres of radius $r = 0.185$~cm at volume filling fraction $\phi = 60$\%. 
In this frequency range, skin depth of aluminum remains on sub-micron level.
This is significantly smaller that particles size in the experiment and it justifies modeling of metal particles as an idealized PEC material. 
A representative structure is shown in Fig.~\ref{fig:simulation_geometry}A.
Spatial discretization employs uniform grid spacing $\Delta = \lambda_0/20 = 0.68$~mm, and perfectly matched layer boundary conditions are applied in all directions. 
A linearly-polarized point source is embedded 3~cm inside the medium within a 3~mm radius air void to prevent placing it inside PEC material, as shown schematically in Fig.~\ref{fig:simulation_geometry}C.
For comparison we also show the geometry where the system is excited externally, Fig.~\ref{fig:simulation_geometry}B. 
As detailed in the following section,
this configuration strongly excites interfacial surface waves and is used here only for comparison.
The detection plane is positioned 1.5~cm from the output surface inside the sample, representing the submerged receiving horn of the experiment. {In the experiment, the detection plane is actually at the surface since microwaves still propagate through the disordered material inside the horn, but the detection plane inside the sample allows us to further suppress surface wave effects.}

Raw field data exhibit strong intensity fluctuations due to wave interference and hot-spot phenomena in disordered media. 
To match experimental conditions where the horn antenna (of aperture area $\sim\lambda_0^2$) spatially integrates the field, we apply spatial convolution with a Gaussian function of full-width-at-half-maximum (FWHM)~$= 2.4$~cm to the intensity distribution $I(x,y,z = L-1.5~\text{cm},t)$. 
Field values in with PEC regions are zero before smoothing. 
The smoothed intensity $I_{\text{smooth}}(x,y,t)$ is azimuthally averaged about the beam axis to obtain $I_{\text{smooth}}(\rho,t)$,
and used to compute $w^2(t)$ from the 2D participation ratio, as explained in the main text.

Results are ensemble-averaged over 10 independent disorder realizations for $L =5$, 6, 8~cm and 20 realizations for $L = 10$, 12~cm, each realization obtained by independent random sphere placement.
Convergence and validation of the FDTD approach for PEC composites in the present parameter regime are established in Ref.~\cite{yamilovAndersonLocalizationElectromagnetic2023}.

\subsubsection*{Surface wave suppression\label{sec:surface_waves}}
\begin{figure} 
	\centering
    \includegraphics[width=0.66\textwidth]{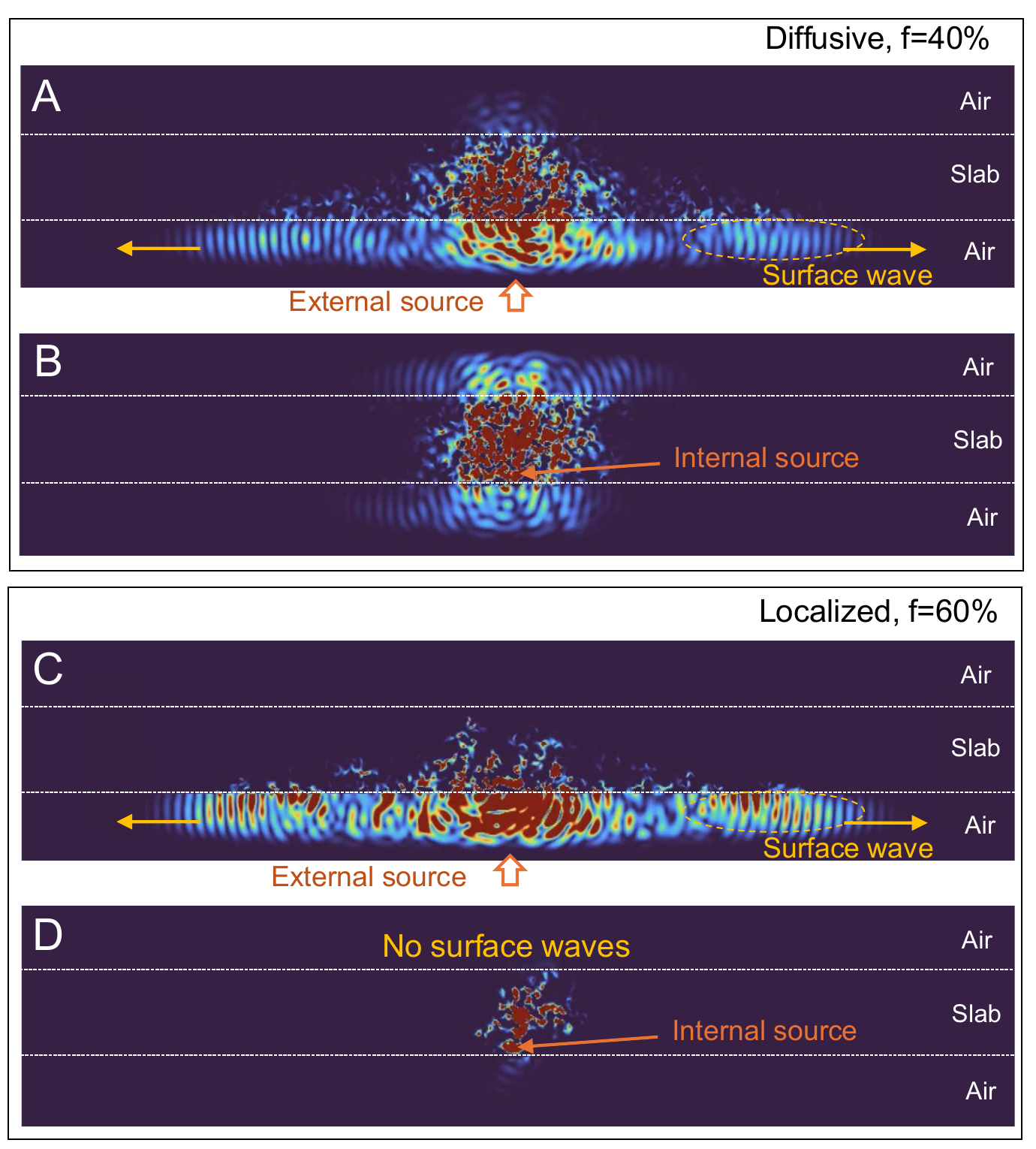}
	\caption{{\textbf{
    Suppression of surface waves by internal excitation.}} 
    Cross-sectional field-intensity maps obtained from FDTD simulations of disordered PEC slabs, showing the effect of excitation geometry in wave transport. Snapshots are obtained at $t = 1$~ns after arrival of main pulse. 
    \textbf{(A)} Diffusive sample ($\phi = 40$\%) under external excitation: pronounced surface waves propagate along both air–slab interfaces over distances well beyond the slab thickness, effectively increasing the source aperture. 
    \textbf{(B)} Same diffusive sample under internal excitation: the intensity distribution is confined to the bulk, with no discernible surface-wave contribution. 
    \textbf{(C)} Localized sample ($\phi = 60$\%) under external excitation: surface waves of amplitude comparable to the bulk field dominate the interfacial region and would mar the transverse intensity profile from bulk transport. \textbf{(D)} Same localized sample under internal excitation: the field remains tightly confined around the source, and surface waves are absent at both interfaces. These results justify the embedded source–detector configuration used both in the simulations and in the experiment reported in the main text.}
	\label{fig:surface_waves} 
\end{figure}

Embedding both source and detector within the disordered medium is essential to suppress surface wave contributions that would otherwise dominate the measured transverse intensity profile. 
When electromagnetic waves are launched from an external source onto a random metal composite, strong scattering at the irregular air-sample interface generates surface waves that propagate along the boundary over distances much larger than the slab thickness. 
These surface modes effectively increase the source aperture, causing the measured beam width to reflect the lateral propagation of the interfacial waves~\cite{2004_Pendry} rather than the transverse extent of excited bulk states.
Because the surface-wave contribution is coherent with the bulk field, it cannot be removed by simple subtraction.
This mechanism is illustrated by FDTD simulations showing field distribution over a cross-sectional cut through the slab, cf. Fig.~\ref{fig:surface_waves}, with snapshots taken 1~ns after the arrival of the main pulse.
For the diffusive sample ($\phi =40$\%), external excitation (Fig.~\ref{fig:surface_waves}A) produces a pronounced wave packet propagating along both air--composite interfaces (clearly visible on the front surface), whereas internal excitation of the same sample (Fig.~\ref{fig:surface_waves}B) yields a compact lateral intensity distribution inside the bulk with no discernible interfacial component. 
The contrast is even more striking in the localized sample ($\phi = 60$\%): external excitation (Fig.~\ref{fig:surface_waves}C) generates surface waves of amplitude comparable to the bulk response, so that the measured transverse profile would be governed almost entirely by interfacial wave propagation rather than by bulk localization, whereas internal excitation (Fig.~\ref{fig:surface_waves}D) produces a tightly confined intensity distribution around the source, with no appreciable surface-wave component at either interface.
By placing the source 3~cm inside the medium (both in the simulations and in the experiment) and recording the transmitted field 1.5~cm inside the sample (in the simulations) or inserting the horn antenna 1.5~cm inside the sample (in the experiment), we therefore selectively excite and probe bulk transport modes while minimizing coupling to interfacial modes. 
This embedded configuration ensures that the spatial scaling of the transverse intensity profile reflects genuine 3D bulk Anderson localization rather than surface wave transport. 
A comprehensive study of the surface-wave phenomena in disordered metallic slabs, including their dispersion and penetration depth, will be presented separately.

\newpage


\end{document}